\documentclass[sigconf]{acmart}

\copyrightyear{2026}
\acmYear{2026}
\setcopyright{cc}
\setcctype{by}
\acmConference[ICSE-Companion '26]{2026 IEEE/ACM 48th International Conference on Software Engineering}{April 12--18, 2026}{Rio de Janeiro, Brazil}
\acmBooktitle{2026 IEEE/ACM 48th International Conference on Software Engineering (ICSE-Companion '26), April 12--18, 2026, Rio de Janeiro, Brazil}
\acmPrice{}
\acmDOI{10.1145/3774748.3795666}
\acmISBN{979-8-4007-2296-7/2026/04}

\def\BibTeX{{\rm B\kern-.05em{\sc i\kern-.025em b}\kern-.08em
    T\kern-.1667em\lower.7ex\hbox{E}\kern-.125emX}}

\usepackage{multirow}
\usepackage{makecell}
\usepackage{tabularx}

\begin{document}

\title{Test-Oriented Programming: rethinking coding for the GenAI era}

\author{Jorge Melegati}
\orcid{https://orcid.org/0000-0003-1303-4173}
\email{melegati@fe.up.pt}
\affiliation{%
  \department{INESC TEC, Faculty of Engineering}
  \institution{University of Porto}
  \city{Porto}
  \country{Portugal}
}


\begin{abstract}
  Large language models (LLMs) have shown astonishing capability of generating software code, leading to its use to support developers in programming. Proposed tools have relied either on assistants for improved auto-complete or multi-agents, in which different model instances are orchestrated to perform parts of a problem to reach a complete solution. We argue that LLMs can enable a higher-level of abstraction, a new paradigm we called Test-Oriented Programming (TOP). Within this paradigm, developers only have to check test code generated based on natural language specifications, rather than focusing on production code, which could be delegated to the LLMs. 
  To evaluate the feasibility of this proposal, we developed a proof-of-concept tool and used it to generate a small command-line program employing two different LLMs. We obtained promising results and identified challenges for the use of this paradigm for real projects. 
\end{abstract}




\maketitle

\section{Introduction}

The emergence of more powerful large language models (LLMs) employed for Generative AI (GenAI) promises to increase developers' productivity and to improve the quality of produced software~\cite{Fan2023}. 
To achieve this aim, current approaches include auto-complete assistants, such as Github Copilot, vibe coding, or the use of multiple agents. However, these approaches still require the inspection of production code by developers which, despite the increased speed of code generation, do not represent an increased level of abstraction.

We argue that GenAI allows the increase of the abstraction level of software development, i.e., instructions could be given to the machine in a language closer to natural languages. This shift is similar to the emergence of high-level languages, e.g., C, that increased the abstraction level of assembly instructions. To this aim, we propose \textbf{Test-Oriented Programming (TOP)}, a novel paradigm consisting of the complete delegation of code generation based on natural language specifications, in which developers are only responsible for verifying test code to solve the ambiguity issue of natural languages.
To assess the feasibility of our proposal, we developed Onion, a proof-of-concept tool that implements systems in Python based on configuration files. Using Onion, a developer only needs to checks and, if needed, modifies configuration files and test code. We evaluated the capacity of the tool for generating a proposed system when relying on different LLM models. In this process, we also identified challenges to be overcome to take this paradigm to real projects. 

\section{Test-Oriented Programming}

Current research and practice have explored the use of GenAI for SE. Initially, coding assistants, e.g., GitHub Copilot, were integrated in IDEs to generate snippets of code, complete functions or classes based on natural-language descriptions or function signatures. 
More recently, multiple agents~\cite{He2025}, i.e., different instances of LLMs, perform specific tasks towards a common goal. 
Finally, the term ``vibe coding'' represents a development workflow based on a conversational process between the developer and a chatbot.
However, these proposals do not change the level of abstraction of programming since the output is still in conventional programming languages to be checked by developers.

To reach higher levels of abstraction, a reasonable candidate would be natural languages, which, however, suffer from ambiguities. 
In conventional programming languages, these ambiguities are solved by formal specifications.
Another issue is the lack of determinism of LLMs, for which a mitigation strategy being explored is the Assured LLM-based Software Engineering~\cite{Alshahwan2024}. 
This approach consists of the application of LLMs to SE in which all LLM responses come ``with some verifiable claim to its utility''~\cite{Alshahwan2024}. For example, test-driven development (TDD) can be employed to guarantee that there exists a way to verify the generated code~\cite{Mock2025}.

We argue that the combination of LLMs and TDD represents a higher-level programming paradigm, which we call Test-Oriented Programming (TOP). TOP is based on creating code for automatic tests (test code), which, alongside natural language specifications, is used to automatically generate production code. It represents an increased abstraction level of programming by completely automatizing the creation of production code, which was previously performed by developers. 
TOP is related to several existent concepts in SE, therefore it is essential to discuss its novelty. 
First, even though TOP is also based on the idea of having tests created before production code as TDD, it proposes another level of programming abstraction which is not the case for TDD in which tests and production code stay at the same level. This shift is only possible thanks to the automatic generation of code enabled by GenAI. 
Second, TOP aims to be a generic paradigm suitable to any problem, requiring trained developers capable of inspecting test code and, as such, is not equivalent of low-code development tools, that are adequate for citizen developers but tailored for specific contexts. 

\section{Preliminary results}

To evaluate the feasibility and identify potential challenges to implement TOP, we developed a proof-of-concept tool called Onion. It is an iterative, command-line tool that, based on a configuration file following a pre-defined template in YAML, generates code by interacting with a LLM with pre-defined prompts and with the developer, responsible for verifying the test code.  
The configuration file contains, as lists of natural language statements, i) a description of the project describing several details of system, including its goal, library dependencies, and outputs and ii) the description of acceptance tests. Based on this information, the tool generates a structure file, also in YAML, containing the description of all the packages to be created and their classes and respective methods. This file could be changed by the developer. 
Then, the tool creates a file containing the acceptance tests and, for each class, a file containing tests. The developer verifies these files and, if needed, modify them. The execution of the tests are then used to generate the code of the classes. If, after a defined number of tries, the generation fails, the tool aborts and the developer should modify the tests, the structure or even the main configuration file. Once all the classes' tests pass, the acceptance tests are run and used to generate the main file of the system. The prompts used in the process were defined through an iterative process to allow the completion of the task when using OpenAI's GPT-4o-mini. The tool is available here\footnote{\url{https://github.com/TOProgramming/onion}}.

To evaluate the tool, we employed it to generate a command-line tool to handle \BibTeX entries, persisting them into a pre-defined file, with options to add entries, list all the entries and search for specific text in the entries.
We employed two different LLMs: OpenAI's GPT-4o mini and Google's Gemini 2.5-Flash, to compare a reasoning and a non-reasoning model.

First, we used the tool to generate a structure configuration file containing all the classes and the respective methods, which we accepted without changes.
Then, keeping the structure unchanged, we generated the production code five times for each LLM from scratch, interacting with the tool as needed. For each run, we collected the interactions needed to complete the tasks, the generated code and all the logs, including the prompts sent to the models. All this data is available as a supplemental package\footnote{\url{https://doi.org/10.5281/zenodo.17227298}}.
 
All the tries were successful but with different levels of developer intervention and code quality. In none of the tries, we had to directly modify the production code. In most of the tries, the generation of the production code failed because of inconsistencies in the test code. Once we fixed these issues, the production code was correctly implemented. However, in two tries, one for each model, we had to add comments in the test code to guide the generation of the production code since the models were always creating code which failed the same tests. 
For GPT-4o-mini, just in one try, we had to modify the test code before checking the implementation. This issue happened more often with the Gemini 2.5-Flash model, however, these issues were more related to supporting code, such as class imports or test execution. A possible explanation for these issues is the fact that prompts had been optimized for another model. 

The speed and amount of test code, especially for the Gemini 2.5-Flash model, represent a challenge for the developer responsible for checking it. In some tries, the fact that the model was not able to generate code that fulfilled the tests prompted the developer to check the test code, identifying issues on it and fixing it. This issue indicates the need for automatic tools to support the developer in the verification of the test code as summarized in the following challenge: \textbf{the amount of the generated test code represents a challenge for developers verifying it.} 
We also observed the issue of the lack of determinism of LLMs in the generation of production and test code: most of the tries led to the generation of different outputs. In summary, \textbf{for the same model, there is some variability of the generated code.} 
Finally, we observed a large difference of the generated code: the ones created by GPT-4o-mini were shorter and had less comments, while the ones created by Gemini 2.5-Flash were longer and contained several comments. 
This issue can be related to the increased verbosity observed in reasoning models~\cite{Jang2025}. This observation leads us to the following challenge: \textbf{there is a huge difference to the generated code for different models.}

\section{Discussion and Conclusion}

Onion is a proof-of-concept and, at the current stage of development, not adequate for the development of a real, complex system. 
We expect that more complex systems would need to be broken into modules, which in their turn could be generated by the tool. For example, in a microservices architecture, each microservice could be a Onion project. However, our evaluation already brings pieces of evidence to support the feasibility of TOP: i) the automatic generation of production code based on test code without human intervention, and ii) of test code based on specifications with minimal human intervention. We also identified three challenges that need to be handled to employ TOP in real projects.


\bibliographystyle{ACM-Reference-Format}
\bibliography{refs}

@inproceedings{Alshahwan2024,
    address = {New York, NY, USA},
    author = {Alshahwan, Nadia and Harman, Mark and Harper, Inna and Marginean, Alexandru and Sengupta, Shubho and Wang, Eddy},
    booktitle = {Proceedings of the ACM/IEEE 2nd International Workshop on Interpretability, Robustness, and Benchmarking in Neural Software Engineering},
    IGNOREdoi = {10.1145/3643661.3643953},
    isbn = {9798400705649},
    month = {apr},
    number = {1},
    pages = {7--12},
    publisher = {ACM},
    title = {{Assured Offline LLM-Based Software Engineering}},
    volume = {1},
    year = {2024}
}

@inproceedings{Fan2023,
    author = {Fan, Angela and Gokkaya, Beliz and Harman, Mark and Lyubarskiy, Mitya and Sengupta, Shubho and Yoo, Shin and Zhang, Jie M.},
    booktitle = {2023 IEEE/ACM International Conference on Software Engineering: Future of Software Engineering (ICSE-FoSE)},
    IGNOREdoi = {10.1109/ICSE-FoSE59343.2023.00008},
    isbn = {979-8-3503-2496-9},
    month = {may},
    pages = {31--53},
    publisher = {IEEE},
    title = {{Large Language Models for Software Engineering: Survey and Open Problems}},
    year = {2023}
}

@article{He2025,
    author = {He, Junda and Treude, Christoph and Lo, David},
    IGNOREdoi = {10.1145/3712003},
    issn = {1049-331X},
    journal = {ACM Transactions on Software Engineering and Methodology},
    month = {jun},
    number = {5},
    pages = {1--30},
    title = {{LLM-Based Multi-Agent Systems for Software Engineering: Literature Review, Vision, and the Road Ahead}},
    volume = {34},
    year = {2025}
}

@inproceedings{Jang2025,
    address = {Stroudsburg, PA, USA},
    author = {Jang, Joonwon and Kim, Jaehee and Kweon, Wonbin and Lee, Seonghyeon and Yu, Hwanjo},
    booktitle = {Findings of the Association for Computational Linguistics: ACL 2025},
    IGNOREdoi = {10.18653/v1/2025.findings-acl.1068},
    number = {Table 1},
    pages = {20769--20784},
    publisher = {Association for Computational Linguistics},
    title = {{Verbosity-Aware Rationale Reduction: Sentence-Level Rationale Reduction for Efficient and Effective Reasoning}},
    year = {2025}
}

@InProceedings{Mock2025,
    author="Mock, Moritz
    and Melegati, Jorge
    and Russo, Barbara",
    IGNOREeditor="Marchesi, Lodovica
    and Goldman, Alfredo
    and Lunesu, Maria Ilaria
    and Przyby{\l}ek, Adam
    and Aguiar, Ademar
    and Morgan, Lorraine
    and Wang, Xiaofeng
    and Pinna, Andrea",
    title="Generative AI for Test Driven Development: Preliminary Results",
    booktitle="Agile Processes in Software Engineering and Extreme Programming -- Workshops",
    year="2025",
    publisher="Springer Nature Switzerland",
    address="Cham",
    pages="24--32",
    isbn="978-3-031-72781-8"
}



\end{document}